\documentclass[pra,aps,amsfonts,amsmath,twocolumn,showpacs,floatfix,superscriptaddress,nofootinbib]{revtex4}
\usepackage{bm}
\usepackage{amsfonts}
\usepackage{amsmath}
\usepackage{epsfig}
\usepackage{epstopdf}
\newcommand{\vek}[1]{\bm{\mathrm{#1}}}
\newcommand{\up}{\uparrow}
\newcommand{\down}{\downarrow}
\newcommand{\Fig}[1]{Fig.~\ref{#1}}
\newcommand{\Sec}[1]{Sec.~\ref{#1}}
\newcommand{\Figs}[1]{Figs.~\ref{#1}}
\newcommand{\Ref}[1]{Ref.~\cite{#1}}
\newcommand{\Refs}[1]{Refs.~\cite{#1}}
\newcommand{\Eq}[1]{Eq.~(\ref{#1})}

\newcommand{\kv}{\vek{k}}
\newcommand{\qv}{\vek{q}}

\newcommand{\NSR}{\text{NSR}}
\newcommand{\BCS}{\text{BCS}}

\DeclareMathOperator{\Imag}{Im}
\DeclareMathOperator{\Real}{Re}

\begin{document}
\title{Polarized Fermi gases at finite temperature in the BCS-BEC crossover}
%
\author{Pierre-Alexandre Pantel}
\affiliation{Universit{\'e} de Lyon, Univ. Lyon 1, CNRS/IN2P3,
IPN Lyon, F-69622 Villeurbanne Cedex, France}
\author{Dany Davesne}
\affiliation{Universit{\'e} de Lyon, Univ. Lyon 1, CNRS/IN2P3,
IPN Lyon, F-69622 Villeurbanne Cedex, France}
\author{Michael Urban}
\affiliation{Institut de Physique Nucl{\'e}aire, CNRS-IN2P3 and
Universit\'e Paris-Sud, 91406 Orsay Cedex, France}
\begin{abstract}
We consider a polarized Fermi gas in the BCS-BEC crossover region
above the critical temperature within a T matrix formalism. By
treating the mean-field like shift of the quasiparticle energies in a
self-consistent manner, we avoid the known pathological behavior of
the standard Nozi\`eres-Schmitt-Rink approach in the polarized case,
i.e., the polarization has the right sign and the spin polarizability
is positive. The momentum distributions of the correlated system are
computed and it is shown that, in the zero-temperature limit, they
satisfy the Luttinger theorem. Results for the phase diagram, the spin
susceptibility, and the compressibility are discussed.
\end{abstract}
\maketitle

\section{Introduction}
Initially, the crossover from the weak-coupling (BCS) superfluid phase
to Bose-Einstein condensation (BEC) of molecules was mainly a
theoretical idea \cite{Eagles,Leggett,NSR}. Its experimental
realization \cite{Greiner} with ultracold trapped fermionic alkali
atoms, whose interaction strength (characterized by the scattering
length $a$) can be tuned with the help of Feshbach resonances,
triggered a lot of activity from the theoretical side, not only in the
context of cold atoms, but also in condensed matter and nuclear
physics \cite{JinUrban,Ramanan}. While at zero temperature the
mean-field (BCS) theory is believed to be reasonable throughout the
crossover (it reproduces the correct wave function of the dimers in
the BEC limit), it fails to describe the critical temperature $T_c$:
on the BEC side, $T_c$ is not the temperature where pairs are formed,
but the temperature where the ``preformed pairs'' condense
\cite{Eagles,Leggett}. By taking into account correlations above $T_c$
in the calculation of the density, Nozi\`eres and Schmitt-Rink (NSR)
obtained a theory that correctly interpolates between the BCS and BEC
critical temperatures.

The situation becomes more complex if the formation of Cooper pairs is
perturbed by a density or mass imbalance between the particles forming
the pairs. Apart from atomic gases and superconductors in magnetic
fields, such situations may be realized in nuclear matter with
different densities of neutrons and protons \cite{Lombardo,Stein2012} or in
compact stars containing light and strange quarks \cite{Alford}. Such
systems have been extensively studied in the recent years. For
instance, one still hopes to find the Fulde-Ferrel-Larkin-Ovchinnikov
(FFLO) phase with a spatially oscillating order parameter
\cite{Bulgac}, which was predicted theoretically a long time ago
\cite{FuldeFerrel,LarkinOvchinnikov} but which no experiment has seen
so far.

In the present work, we concentrate on the case of a uniform Fermi gas
with two components (denoted by $\sigma = \up,\down$ in analogy with
spin-1/2 systems) having equal masses $m$ but different densities
$\rho_\up\geq \rho_\down$. Although experiments are generally done in
traps, they provide information about the uniform gas under the
assumption of the validity of the local-density approximation
(LDA). The experiments on the phase diagram of polarized\footnote{The
  term ``polarized'' refers to a finite polarization (density
  imbalance) and does not imply that the gas is fully polarized.}
Fermi gases done at MIT \cite{Zwierlein2006,Shin2006,Shin2008} and
Paris \cite{Nascimbene} can be well understood within the LDA. Only in
an experiment at Rice University \cite{Partridge}, clear deviations
from LDA were observed, but they were later shown to correspond to a
metastable non-equilibrium configuration of the atomic cloud
\cite{ParishHuse2009,Liao2011}.

From the theoretical side, different approaches were used to describe
the finite-temperature phase diagram of the polarized Fermi gas. Let
us mention the quantum Monte-Carlo (QMC) calculations of
\Ref{Pilati2008} and the Wilsonian renormalization group approach of
\Ref{Gubbels2008}. Also the NSR approach mentioned above was
generalized to the polarized case \cite{LiuHu06}. However, it turned
out that it breaks down near the unitary limit ($a\to\infty$): one
finds that the sign of the polarization is opposite to that of the
difference between the two chemical potentials
\cite{LiuHu06,ParishMarchetti07}. Actually, there is already a problem
in the unpolarized case, where the NSR approach gives a negative spin
susceptibility $\chi$ \cite{Kashimura2012}. This is surprising, since
a very similar approach, also based on the T matrix in ladder
approximation, works very well in the extremely polarized limit at
zero temperature (``polaron'') \cite{Combescot2007}. Some modified
versions of the NSR scheme have been developed that avoid the
unphysical behavior, such as the ``extended T-matrix approximation''
(ETMA) by Kashimura et al. \cite{Kashimura2012} or the $GG_0$ approach
by Chen et al. \cite{Chen2005}, which was also applied to the
polarized case \cite{ChenHe2007,HeChien2007}. Roughly speaking, these
modified versions of the NSR scheme are based on dressing a propagator
line in the Feynman diagrams for the self-energy: in the ETMA, it is
the upper line in \Fig{fig:feynman}b, while in the $GG_0$ approach, it
is one of the two lines in the ladder diagrams of \Fig{fig:feynman}a.

The goal of the present work is to develop a scheme similar to the NSR
approach which allows us to describe the polarized Fermi gas in the
normal-fluid phase from the unpolarized case above $T_c$ up to the
polaron limit. We will see that the problem of the NSR approach is
caused by its non-selfconsistent treatment of the quasiparticle energy
shift generated by the self-energy. By simply including this shift
self-consistently in \emph{all} lines of the Feynman diagrams, the
problem of the unphysical sign of the polarization is avoided. Within
this framework, we will compute the phase transition line towards the
paired (superfluid or FFLO) phase, the equation of state in the normal
phase, and the correlated occupation numbers.

The article is organized as follows. We present the formalism in
\Sec{sec:formalism}. Then we discuss the correlated density and
occupation numbers (\Sec{sec:density}), the phase diagram
(\Sec{sec:phasediagram}) and the compressibility and spin
susceptibility (\Sec{sec:susceptibility}). Finally,
\Sec{sec:conclusions} contains the summary and further
discussions. Throughout the article, we use units with $\hbar = k_B =
1$, where $\hbar$ and $k_B$ denote, respectively, the reduced Planck
constant and the Boltzmann constant.
\section{Formalism}\label{sec:formalism}
\begin{figure}
\epsfig{file=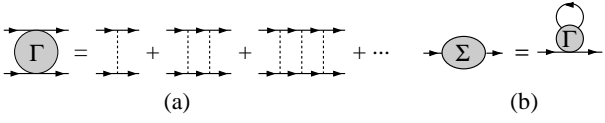,width=8cm}
\caption{Ladder diagrams for the in-medium T matrix (a) and the
  self-energy (b).\label{fig:feynman}}
\end{figure}

The starting point of the NSR approach and most of its variants is the
T matrix, i.e., the interaction in the medium is calculated in ladder
approximation (see \Fig{fig:feynman}a)
\begin{equation}
\Gamma(\Omega,\qv) = \left(\frac{m}{4\pi a}-J(\Omega,\qv)\right)^{-1}
\end{equation}
where $\Omega$ and $\qv$ are energy (measured relatively to the sum of
the two chemical potentials, $\mu_\up+\mu_\down$) and momentum of the
pair,
\begin{equation}
J(\Omega,\qv) = \int \frac{d^3k}{(2\pi)^3}
  \left(\frac{1-f(\xi^*_{\kv+\qv,\up})-f(\xi^*_{\kv,\down})}
    {\Omega-\xi^*_{\kv+\qv,\up}-\xi^*_{\kv,\down}}+\frac{m}{k^2}\right)
\label{J}
\end{equation}
is the retarded in-medium two-particle propagator (regularized in the
usual way \cite{SadeMelo}), $\xi^*_{\kv\sigma}$ are the quasiparticle
energies (measured with respect to $\mu_\sigma$), and $f(\xi) =
1/(e^{\xi/T}+1)$ is the Fermi function for temperature $T$. In the
standard NSR theory \cite{NSR,SadeMelo}, one takes instead of
$\xi^*_{\kv\sigma}$ the free-particle dispersion relation,
$\xi^0_{\kv\sigma} = k^2/(2m)-\mu_\sigma$. In the present work, we
will make a step towards a more self-consistent treatment by using a
modified dispersion relation that includes interaction effects. In
principle, it would be desirable to calculate $\xi^*_{\kv\sigma}$ by
looking for the pole of the dressed Green's function, i.e., from
\begin{equation}
\xi^*_{\kv\sigma} = \xi^0_{\kv\sigma} + \Real \Sigma_\sigma(\xi^*_{\kv\sigma},\kv)\,.
\end{equation}
In ladder approximation, the self-energy $\Sigma_\sigma$ of a particle
with spin $\sigma$ is calculated from $\Gamma$ by closing the line
corresponding to the particle with the opposite spin, $\bar{\sigma}$,
as shown in \Fig{fig:feynman}b. Calculating this diagram within the
imaginary-time (Matsubara) formalism and performing the analytic
continuation to real energies \cite{FetterWalecka} one obtains for the
imaginary part of the retarded self-energy
\begin{multline}
\Imag \Sigma_\sigma(\omega,\kv) = -\int\frac{d^3k'}{(2\pi)^3}
  \Imag\Gamma(\omega+\xi^*_{\kv'\sigma},\kv+\kv')\\
  \times [f(\xi^*_{\kv'\bar{\sigma}})+g(\omega+\xi^*_{\kv'\sigma})]\,,
\end{multline}
where $g(\omega) = 1/(e^{\omega/T}-1)$ denotes the Bose function. The
real part can be obtained from the imaginary part with the help of a
dispersion relation,
\begin{equation}
\Real \Sigma_\sigma(\omega,\kv) = -\mathcal{P}\int
  \frac{d\omega'}{\pi} \frac{\Imag\Sigma_\sigma(\omega',\kv)}{\omega-\omega'}\,.
\end{equation}

Within the original NSR theory \cite{NSR,SadeMelo}, the density is
obtained from the thermodynamic potential in ladder
approximation. This is equivalent to calculating the density from the
Green's function obtained by truncating the Dyson equation at first
order, i.e., $G_\sigma = G^0_\sigma + G^{0\,2}_\sigma \Sigma_\sigma$
\cite{Chen2005}, where $G^0_\sigma = 1/(\omega-\xi^0_{\kv\sigma})$
denotes the non-interacting Green's function. As a consequence, the
density for each spin state has two contributions,
$\rho^{\phantom{(}}_\sigma = \rho^{(0)}_\sigma +
\rho^{(1)}_\sigma$, where $\rho^{(0)}_\sigma$ is the density of
an ideal Fermi gas with chemical potential $\mu_\sigma$, and the
correction $\rho^{(1)}_\sigma$ is given by
\begin{equation}
\rho^{(1)}_\sigma =
  \frac{\partial}{\partial\mu_\sigma}\int \frac{d^3q}{(2\pi)^3}\int
  \frac{d\Omega}{\pi} g(\Omega) \delta(\Omega,\qv)\,,
  \label{rhocorrnsr}
\end{equation}
where
\begin{equation}
\delta(\Omega,\qv) = -\Imag \ln\Big(J(\Omega,\qv)-\frac{m}{4\pi a}\Big)
\end{equation}
is the in-medium scattering phase shift. In the presence of a bound
state ($a > 0$), one has $\delta = \pi$ in the energy range between
the bound-state energy and the continuum threshold.

In other variants of the NSR theory, the Dyson series has been
resummed to all orders, i.e., $G_\sigma =
1/(\omega-\xi^0_{\kv\sigma}-\Sigma_\sigma)$ \cite{Perali2002}. In
either way, $\Sigma_\sigma$ describes at the same time correlation
effects and a mean-field like shift of the single-particle energies.

In the present work, the situation is slightly different. The bare
lines correspond already to quasiparticle Green's functions
$G^*_\sigma = 1/(\omega-\xi^*_{\kv\sigma})$ that contain the modified
dispersion relation $\xi^*_{\kv\sigma}$. The mean-field like shift is
thus included self-consistently (also in the calculation of
$\Gamma$). The additional correlation effects are responsible for the
$\omega$-dependence of $\Sigma_\sigma$. To first order in the
correlations, we therefore get
\begin{multline}
G_\sigma(\omega,\kv) = G^*_\sigma(\omega,\kv)\\ 
  + G^{* 2}_{\sigma}(\omega,\kv) 
  [\Sigma_\sigma(\omega,\kv)-\Real \Sigma_\sigma(\xi^*_{\kv\sigma},\kv)]\,.
\label{Gperturbative}
\end{multline}
Using this approximation, one can express the occupation numbers
$n_{\kv\sigma}$ in the form $n^{\phantom{(}}_{\kv\sigma} = n^{(0)}_{\kv\sigma} +
n^{(1)}_{\kv\sigma}$ as a sum of the uncorrelated occupation numbers
$n^{(0)}_{\kv\sigma} = f(\xi^*_{\kv\sigma})$ and a correction due to
correlations
\begin{equation}
n^{(1)}_{\kv\sigma} = \int \frac{d\omega}{\pi}\Imag \Sigma_\sigma(\omega,\kv)
  \frac{f(\xi^*_{\kv\sigma})-f(\omega)}{(\omega-\xi^*_{\kv\sigma})^2}\,.
\end{equation}
Accordingly, the densities are again a sum of uncorrelated and
correlated densities, $\rho^{\phantom{(}}_\sigma = \rho^{(0)}_\sigma +
\rho^{(1)}_{\phantom{\sigma}}$, but now $\rho^{(0)}_\sigma$ is
the density of an uncorrelated gas of quasiparticles with dispersion
relation $\xi^*_{\kv\sigma}$. After some algebra, the expression for
the correction $\rho^{(1)}$ can be reduced to
\begin{equation}
\rho^{(1)} = -\int \frac{d^3q}{(2\pi)^3}\int \frac{d\Omega}{\pi}
   g'(\Omega)
   \left(\delta - \frac{1}{2}\sin(2\delta)\right)\,.
  \label{rhocorrzs}
\end{equation}
Note that, in contrast to \Eq{rhocorrnsr}, in the present approach the
correlated density $\rho^{(1)}$ is independent of the spin
$\sigma$. This is plausible since a correlated pair consists of one
atom of each spin. The expression (\ref{rhocorrzs}) for the correlated
density was originally derived for the unpolarized case by Zimmermann
and Stolz (ZS) in \Ref{ZimmermannStolz} in a condensed-matter context
and subsequently used in \Refs{SchmidtRoepke,SteinSchnell,JinUrban} to
describe the BEC-BCS crossover in nuclear matter.

For practical reasons, in order to simplify the numerical
calculations, we make an additional approximation: we replace the
momentum dependent shift $\Real \Sigma_\sigma(\xi^*_{\kv\sigma},\kv)$
in the quasiparticle energies $\xi^*_{\kv\sigma}$ by a constant shift
$U_\sigma$, calculated at the respective Fermi surface, $k_{F\sigma} =
(6\pi^2\rho_\sigma)^{1/3}$, i.e., we use
\begin{equation}
\xi^*_{\kv\sigma} \approx \xi^0_{\kv\sigma} +U_\sigma\,,\quad\text{with}\quad 
  U_\sigma = \Real \Sigma_\sigma(\xi^*_{k_{F\sigma}},k_{F\sigma})\,.
\label{eq:constantshift}
\end{equation}
With this approximation, the T matrix $\Gamma$ and the self-energies
$\Sigma_\sigma$ are identical to those of the standard T-matrix
approximation if one replaces the chemical potentials $\mu_\sigma$ by
``effective'' ones
\begin{equation}
\mu^*_\sigma = \mu_\sigma-U_\sigma\,.
\end{equation}
Actually, as long as one is not interested in the ``real'' chemical
potentials $\mu_\sigma$, it is not necessary to compute the shift
$U_\sigma$ and all calculations can be done as functions of
$\mu^*_\sigma$. However, the shift $U_\sigma$ is needed if one is
interested in the relation between the densities $\rho_\sigma$ and the
real chemical potentials $\mu_\sigma$.

\section{Correlated densities and occupation numbers}\label{sec:density}
\begin{figure}
\epsfig{file=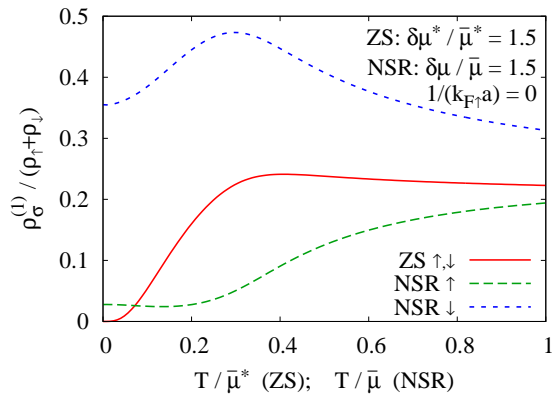,width=7.5cm}
\caption{Temperature dependence of the correlation correction to the
  densities within the present (ZS) approach (solid line) compared
  with the corrections to the majority (long dashes) and minority
  (short dashes) densities within the original NSR approach.
  \label{fig:rhocorrvst}}
\end{figure}

As we discussed before, the main difference between the mean-field
approach on the one hand and the NSR and ZS approaches on the other
hand is the inclusion of pair correlations above $T_c$ in the
calculation of the density. If we consider a sufficiently strong
asymmetry of the densities or chemical potentials, we can calculate
the correlation correction to the density as a function of temperature
down to $T=0$ without ever reaching the superfluid phase. As an
example, we show in \Fig{fig:rhocorrvst}
the temperature dependence of the correction to the density,
$\rho^{(1)}$, obtained within the present approach (ZS) and the
temperature dependence of the corrections $\rho^{(1)}_\sigma$ obtained
within the original NSR approach, normalized to the total density
$\rho_\up+\rho_\down$, in the unitary limit. In the NSR case, we kept
the chemical potentials $\mu_{\up,\down} = \bar{\mu}\pm\delta\mu/2$
constant, with $\delta\mu = 1.5\bar{\mu}$, whereas in the ZS case, we
fixed for simplicity the effective chemical potentials
$\mu^*_{\up,\down}$.

One sees that the behaviors of the spin-independent correction
$\rho^{(1)}$ in the ZS case and of the spin-dependent ones,
$\rho_\sigma^{(1)}$, in the NSR case are qualitatively different:
in the ZS case, the correction vanishes in the limit of zero
temperature. In contrast, in the NSR scheme, the correction
$\rho^{(1)}_\sigma$ does not only account for correlations, but also
for the mean-field like shift of quasiparticle energies (which within
the ZS scheme is included in the effective chemical potentials
$\mu^*_\sigma$). Since the minority atoms ($\down$) feel a much
stronger attractive ``mean field'' than the majority atoms ($\up$),
the NSR correction $\rho^{(1)}_\down$ is much larger than
$\rho^{(1)}_\up$, and both corrections remain finite at $T=0$.

The fact that within the ZS approach $\rho^{(1)}\to 0$ for $T\to 0$ is
directly related to the Luttinger theorem \cite{Luttinger}. This
theorem states that, at $T=0$, the relationship between the density
$\rho_\sigma$ and the Fermi momentum (i.e., the momentum where the
occupation numbers are discontinuous), $\rho_\sigma =
k_{F\sigma}^3/(6\pi^2)$, remains unchanged even though correlations
modify the occupation numbers. Since this relation is already
fulfilled with the uncorrelated occupation numbers
$n^{(0)}_{\kv\sigma} = \theta(k_{F\sigma}-k)$, this implies that the
integral of the correction $n^{(1)}_{\kv\sigma}$, i.e., $\rho^{(1)}$,
must vanish at $T=0$. Our numerical results show that this is indeed
the case. It is interesting to notice that by including the shift
$U_\sigma$ self-consistently, one apparently recovers in the $T\to 0$
limit the results of the $T=0$ formulation of the so-called
``particle-particle random-phase-approximation'', where one starts
from the beginning with Green's function that depend on $k_{F\sigma}$
and not on $\mu_\sigma$ \cite{UrbanSchuck}.

\begin{figure*}
\epsfig{file=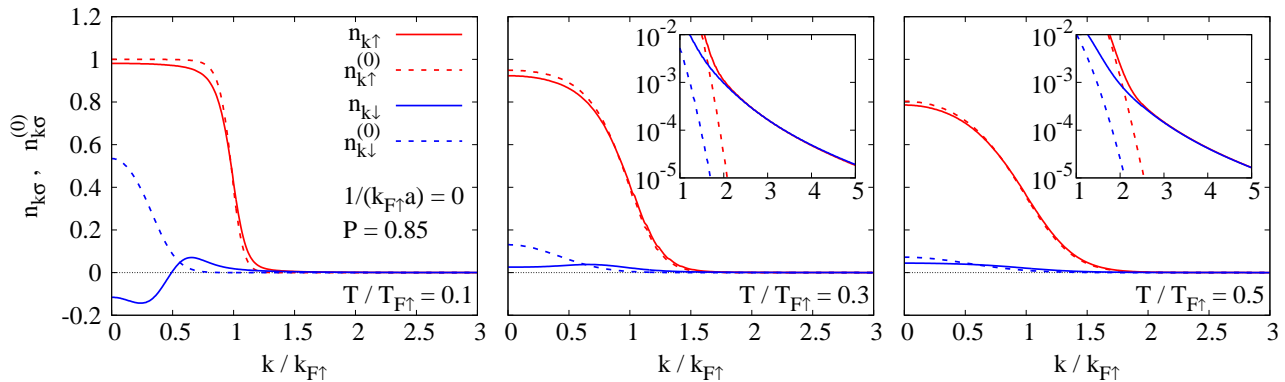,width=17cm}
\caption{Momentum dependence of the occupation numbers in the unitary
  limit for polarization $P=0.85$ and three different temperatures
  $T/T_F = 0.1$, $0.3$, and $0.5$ (from left to right). The insets
  show the asymptotic high-momentum tail of the occupation numbers on
  a logarithmic scale. The upper (red) curves represent the occupation
  numbers of the majority ($\up$) atoms, while the lower (blue) lines
  represent those of the minority ($\down$) atoms. The solid lines are
  the correlated occupation numbers $n_{\kv\sigma}$ while the dashed
  lines represent the uncorrelated ones $n^{(0)}_{\kv\sigma}$.
\label{fig:occ}}
\end{figure*}

Let us now have a look at the occupation numbers themselves. In
\Fig{fig:occ}
we show the occupation numbers for $\up$ and $\down$ atoms in the
unitary limit for a given polarization $P =
(\rho_\up-\rho_\down)/(\rho_\up+\rho_\down) = 0.85$ for three
different temperatures. Let us first discuss the case of $T = 0.5 T_F$
(right panel). At first glance, the change between uncorrelated
($n^{(0)}_{\kv\sigma}$) and correlated ($n_{\kv\sigma}$) occupation
numbers seems to be very small in this case. One sees that
$n_{\kv\sigma}$ is slightly reduced at small $k$ and slightly enhanced
at large $k$ as compared with $n^{(0)}_{\kv\sigma}$. However, the main
effect becomes visible if we look at the asymptotic high-momentum tail
shown in the inset. The correlated occupation numbers
($n_{\kv\sigma}$) fall off like $1/k^4$, the coefficient being the
same for both spins, while the uncorrelated ones
($n^{(0)}_{\kv\sigma}$) decrease of course exponentially.

At a lower temperature, $T = 0.3 T_F$ (central panel), we see in
addition to the increase of the correlated occupation numbers at high
momenta a sizeable reduction at low momenta, especially for the
minority component. This fits into the common picture of how
correlations modify the occupation numbers at $T=0$: particles are
scattered out of the Fermi sea, which reduces the occupation numbers
below $k_{F\sigma}$ and leads to a finite occupancy of states above
$k_{F\sigma}$. In the present case, the Fermi surfaces are of course
washed out by the finite temperature.

As it was already pointed out in \Ref{UrbanSchuck}, the fact that the
pairs are always formed of one $\up$ and one $\down$ atom implies
that, at $T=0$, the depletion of the particle number inside the Fermi
sphere is the same for both spins, i.e., the occupation numbers of the
minority ($\down$) species are necessarily more strongly reduced than
those of the majority species ($\up$). As we see, this effect persists
at finite temperature.

At even lower temperature, $T = 0.1 T_F$ (left panel), we observe that
the correlation correction to the minority occupation numbers becomes
so strong that the occupation numbers $n_{\kv\down}$ become
negative. This is of course unphysical and shows the limits of the
perturbative treatment of the correlations, i.e., of the truncation of
the Dyson series at first order in the self-energy in
\Eq{Gperturbative}. The same problem was found in
\Ref{UrbanSchuck}. However, as soon as one goes more towards the BCS
side of the crossover, this problem appears only at very low
temperatures and very close to the critical polarization.

\section{Phase diagram}\label{sec:phasediagram}
\begin{figure}
\epsfig{file=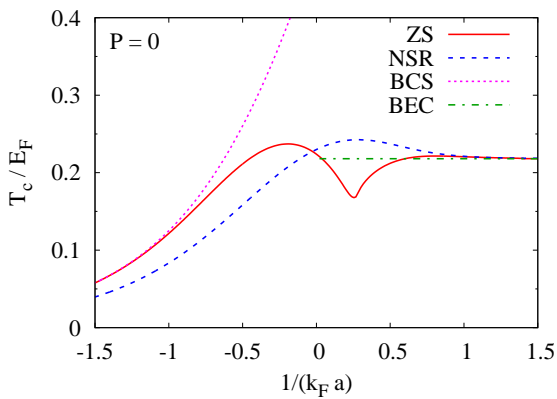,width=7.5cm}
\caption{Critical temperature $T_c$ in units of $E_F$ vs. the
  dimensionless parameter $1/(k_Fa)$ characterizing the interaction
  strength for the unpolarized gas. Solid line: present approach (ZS),
  dashes: NSR result, dots: BCS result, dash-dots: BEC limit.
  \label{fig:bcs_nsr_zs}}
\end{figure}

Before we consider the phase diagram of the polarized gas, let us
briefly discuss the unpolarized case and compare our results with
those of the original NSR theory. As in the NSR approach
\cite{SadeMelo}, the critical temperature is determined from the
Thouless criterion $\Gamma^{-1}(0,0) = 0$, i.e.,
\begin{equation}
J(0,0) = \frac{m}{4\pi a}\,,
\label{thouless}
\end{equation}
but now, $J$ is calculated with the chemical potential
$\mu^*=\mu^*_\up=\mu^*_\down$ (in the unpolarized case we can drop the
spin indices) and the corresponding density is obtained from the ZS
formula (\ref{rhocorrzs}). In \Fig{fig:bcs_nsr_zs},
we display the critical temperature $T_c$ in units of the Fermi energy
$E_F = k_F^2/(2m)$ as function of the dimensionless parameter
$1/(k_Fa)$ characterizing the interaction strength. For the solid
line, $k_F$ was calculated with the density corrected by the ZS
formula (\ref{rhocorrzs}), while the dashed line was obtained with the
standard NSR correlated density \cite{SadeMelo}.

Both theories interpolate between the mean-field (BCS) result
(corresponding to $\rho = \rho^{(0)}$ without correction; dotted line)
in the limit $1/(k_Fa)\to -\infty$ and the condensation temperature
for an ideal gas of bosonic molecules in the limit $1/(k_Fa)\to
\infty$ (BEC, dash-dotted line). However, we see that on the side
$1/(k_Fa) < 0$, the ZS formula reaches the BCS limit much faster than
the NSR one, which gives $T_c^{\NSR} < T_c^{\BCS}$ even for relatively
weak interactions. This reduction of $T_c^{\NSR}$ in the weak-coupling
regime looks similar to the Gor'kov-Melik-Barkhudarov (GMB) correction
\cite{GMB} to $T_c^{\BCS}$, however its origin is completely
different: while the GMB correction is due to screening of the
interaction in the medium, the reduction of $T_c^{\NSR}$ comes from
the non self-consistent treatment of the mean-field like shift in the
original NSR theory.

Coincidentally, the critical temperatures obtained with the ZS and NSR
formulas in the unitary limit ($1/(k_Fa) = 0$) are very close to each
other ($T_c/E_F\approx 0.23$). Although they are much lower than the
BCS result ($T_c^{\BCS}/E_F \approx 0.5$), they are still too high because of
missing screening effects: recent experimental values range from
$T_c/E_F = 0.157(15)$ \cite{Nascimbene} to $0.167(13)$ \cite{Ku2012}.

On the BEC side ($a > 0$), the ZS critical temperature goes through a
minimum before it rises again and approaches the BEC limit, whereas
the NSR critical temperature goes through a maximum. Qualitatively,
the NSR behavior is in better agreement with QMC results
\cite{Pilati2008} than the ZS one. The presence of a minimum in the ZS
critical temperature seems to be a general property of this approach,
cf. the results in the literature for nuclear matter
\cite{SchmidtRoepke,SteinSchnell,JinUrban}.

Let us now turn to the polarized case. Again, the critical temperature
(or polarization) is determined by the appearance of a pole in the T
matrix at $\omega = 0$, but in the polarized case it may happen that
the pole appears first (for decreasing temperature or polarization) at
a finite value of $\qv$, corresponding to the transition to a
FFLO-like phase \cite{LiuHu06}. Therefore the condition to be in the
normal phase reads:
\begin{equation}
J(0,\qv) > \frac{m}{4\pi a}\quad \mbox{for all $\qv$}.
\label{thoulessloff}
\end{equation}

\begin{figure}
\epsfig{file=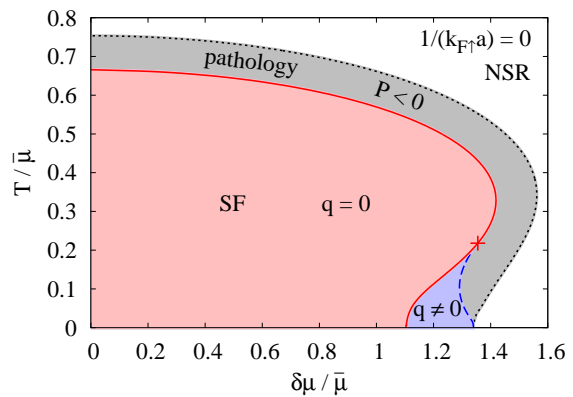,width=7.5cm}
\caption{Phase diagram of the unitary Fermi gas obtained within the
  standard NSR scheme as function of temperature $T$ and difference of
  the chemical potentials $\delta\mu$, both normalized by the average
  chemical potential $\bar{\mu}$. The solid line indicating the
  transition towards a BCS- (or Sarma-) like superfluid (SF) phase is
  obtained with the usual Thouless criterion (\ref{thouless}), whereas
  the dashed line accounts for the possibility of a FFLO-like phase
  with a finite momentum $\qv\ne 0$ of the Cooper pairs,
  cf. \Eq{thoulessloff}. In the region delimited by the dotted line,
  the NSR approach presents a pathology in the sense that the
  polarization has the wrong sign.
  \label{fig:phasediagnsr}}
\end{figure}

As mentioned in the introduction, the standard NSR theory presents in
the polarized case a pathology near the unitary limit: for $\mu_\up >
\mu_\down$, one finds $\rho_\up < \rho_\down$ in large regions of the
phase diagram \cite{LiuHu06,ParishMarchetti07}. To illustrate this
problem we show in \Fig{fig:phasediagnsr}
the phase diagram obtained within the NSR scheme for $1/(k_Fa) =
0$. The pathology is present in the gray shaded region delimited by
the dotted line. Since the pathology extends down to $\delta\mu = 0$,
it implies that the spin susceptibility of the unpolarized gas is
negative in some temperature range above $T_c$
\cite{Kashimura2012}. As we will see, the self-consistent treatment of
the shift $U_\sigma$ in our approach cures these problems.

\begin{figure}
\epsfig{file=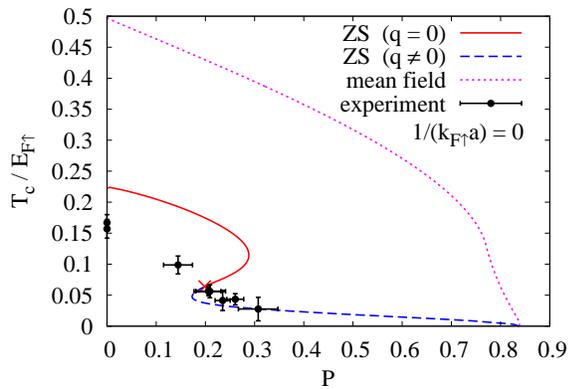,width=7.5cm}
\caption{Critical temperature $T_c$ in units of $E_{F\uparrow} =
  k_{F\uparrow}^2/(2m)$ vs.\@ polarization $P$ for a unitary Fermi gas
  ($1/(k_{F\up}a) = 0$). The transition towards a BCS- (or Sarma-)
  like superfluid is shown as the solid line, while the dashed line
  indicates a transition towards a FFLO-like phase. The cross marks
  the tricritical point separating BCS, FFLO, and normal phase. For
  comparison, the $T_c$ vs.\@ $P$ curve obtained without the
  correction $\rho^{(1)}$ (BCS mean-field result) is shown as the
  dotted line. The experimental data are from Shin et
  al. \cite{Shin2008} except the points at $P=0$ which are from
  Nascimb\`ene et al. \cite{Nascimbene} ($T_c/T_F = 0.157(15)$) and Ku
  et al. \cite{Ku2012} ($T_c/T_F = 0.167(13)$).
  \label{fig:phasediagunitary}}
\end{figure}

\begin{figure}
\epsfig{file=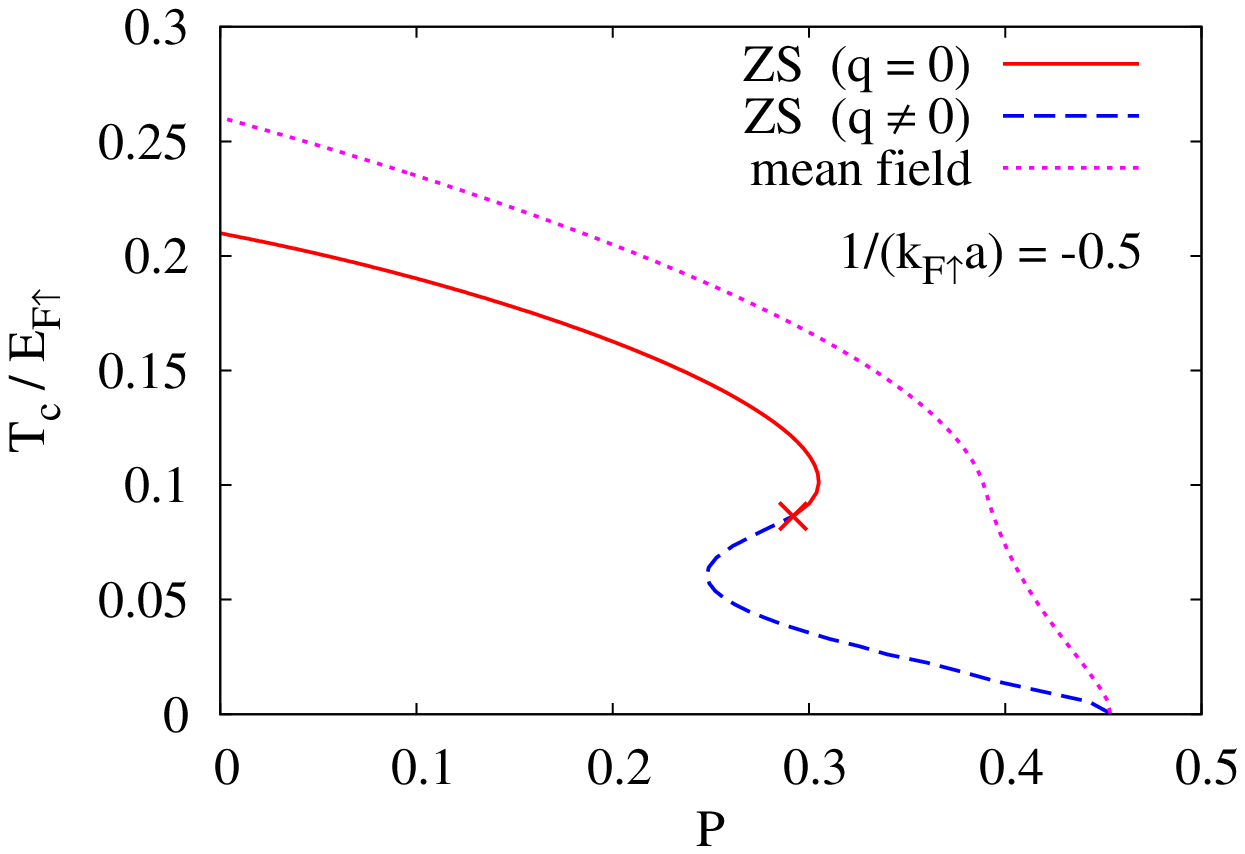,width=7.5cm}
\caption{Same as \Fig{fig:phasediagunitary} but for $1/(k_{F\up}a) =
  -0.5$.
  \label{fig:phasediagbcs}}
\end{figure}

Let us discuss the phase diagram within our approach as function of
temperature $T$ and polarization $P =
(\rho_\up-\rho_\down)/(\rho_\up+\rho_\down)$. The phase diagrams for
two different values of the interaction strength, $1/(k_{F\up}a) = 0$
and $-0.5$, are shown in \Figs{fig:phasediagunitary}
and \ref{fig:phasediagbcs},
respectively. The solid and dashed lines indicate the results obtained
for the critical temperature $T_c$ within the generalized ZS scheme,
while the dotted lines are mean-field results, i.e., what one obtains
if one neglects the correlation contribution $\rho^{(1)}$ to the
density. The cross marks the tricritical point where the phase
transition is not longer towards the ordinary BCS-like superfluid or
Sarma phase ($\qv=0$) but towards a FFLO-like phase ($\qv\neq
0$). Since our formalism does not allow us to calculate the densities
inside the superfluid phase, we cannot draw the line separating these
phases.

In both cases, $1/(k_Fa) = 0$ and $-0.5$, we checked that the difference
between the real chemical potentials, $\delta\mu$, is always positive
for $P>0$, i.e., the pathology of the NSR scheme is not present here.

One sees that in the case $1/(k_Fa)=-0.5$ (\Fig{fig:phasediagbcs}),
the BCS mean-field result clearly differs from that of the full
calculation, but the difference is not dramatic. If one goes further
to the BCS regime [$1/(k_Fa) < -1$], the BCS mean-field and full
calculations give practically identical results. However, as mentioned
before, one should be aware of the fact that BCS mean-field theory as
well as our calculation miss corrections due to screening of the
in-medium interaction. Therefore it seems likely that not only our
critical temperatures, but also the critical polarizations are too
high.

In the unitary limit (\Fig{fig:phasediagunitary}), the inclusion of the
correlated density $\rho^{(1)}$ changes the phase diagram completely,
as expected from our results discussed above for the unpolarized case.
Again, compared with the results of the MIT and ENS experiments
\cite{Shin2008,Ku2012,Nascimbene}, the critical temperature $T_c$ we
obtain at small polarization $P$ is still too high because of missing
screening effects.

In the region of lower temperature and higher polarization ($P \gtrsim
0.2$), the experiment found a first-order phase transition (phase
separation), while we get a second-order phase transition towards the
FFLO phase in this region. We cannot check whether there is a
first-order phase transition since this requires to compare the
energies of the paired and the unpaired phases, the former being
inaccessible within our formalism. But it is clear that, if there was
a first-order phase transition, the critical polarization would have
to be higher than the one where we find the second-order phase
transition.

At very low temperature, the critical polarization increases a lot and
exceeds by far the experimental one. Actually, in our formalism, the
critical polarization beyond which the system stays in the normal
phase even at $T=0$ (Chandrasekhar-Clogston limit) is the same as the
one obtained in mean-field theory. The reason for this is that, as
discussed in the preceding section, the correlated density
$\rho^{(1)}$ vanishes in the $T\to 0$ limit, as required by the
Luttinger theorem. This high value of the critical polarization is
probably due to the uncorrelated occupation numbers
$n^{(0)}_{\kv\sigma}$ in \Eq{J}, which are almost step functions at
low temperature. Maybe a more self-consistent treatment of
correlations, i.e., the inclusion of correlated occupation numbers
$n_{\kv\sigma}$ in \Eq{J}, could improve the results. In nuclear
physics, such an approach is known as ``renormalized random-phase
approximation'' \cite{Catara,Delion}, but it is beyond the scope of
this work.

\section{Spin susceptibility and compressibility}\label{sec:susceptibility}
\begin{figure}
\epsfig{file=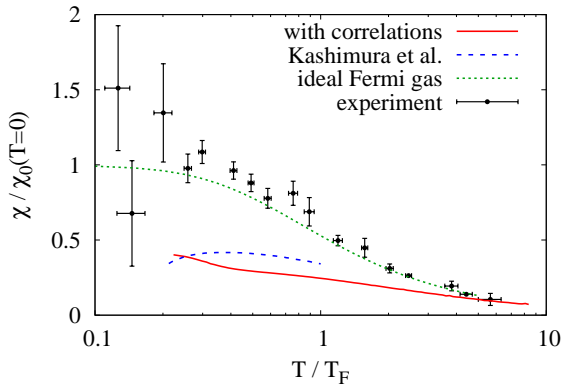,width=7.5cm}
\caption{Spin susceptibility (normalized to the susceptibility of an
  ideal Fermi gas at $T=0$) of the unpolarized unitary Fermi gas as a
  function of temperature. The experimental data are taken from Sommer
  et al. \cite{Sommer}.
  \label{fig:chi}}
\end{figure}

By considering a very small polarization, we can determine the spin
susceptibility of the unpolarized gas. To be precise, the spin
susceptibility is defined as \cite{Kashimura2012}
\begin{equation}
\chi = \lim_{\delta\mu\to 0} \frac{\rho_\up-\rho_\down}{\delta\mu}\,.
\end{equation}
Note that for the computation of $\delta\mu$ (not $\delta\mu^*$) one
needs the self-energy, cf. \Eq{eq:constantshift}. In \Fig{fig:chi},
we show the temperature dependence of the spin susceptibility for
$1/(k_{F\up}a) = 0$ (solid line), in units of the spin susceptibility
of an ideal Fermi gas at zero temperature, $\chi_0(T=0) =
mk_F/(2\pi^2)$. First of all, we see that $\chi$ is positive, which is
already a good point. For comparison, the temperature dependence of
the susceptibility of an ideal Fermi-gas, $\chi_0$, is shown, too
(dotted line). It seems plausible that the susceptibility $\chi$ of
the correlated system is lower than that of the ideal gas, $\chi_0$,
because the pairs made of $\up$ and $\down$ atoms resist against
polarization. Similar results were obtained by Kashimura et
al. \cite{Kashimura2012} within the ETMA (dashes). Very surprisingly,
the experimental results for $\chi$ \cite{Sommer} are close to
$\chi_0$ or even higher. However, one should notice that these data
were determined very indirectly from a complicated non-equilibrium
situation.

\begin{figure}
\epsfig{file=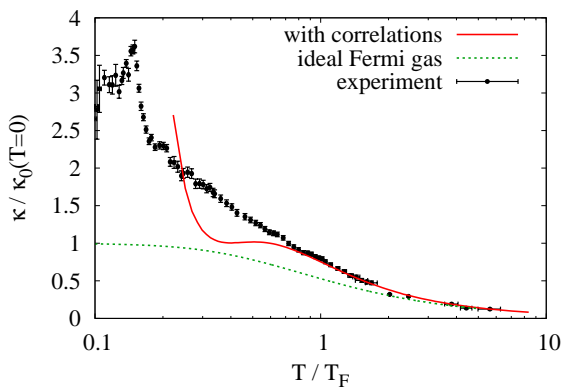,width=7.5cm}
\caption{Compressibility (normalized to the compressibility of an
  ideal Fermi gas at $T=0$) of the unpolarized unitary Fermi gas as a
  function of temperature. The experimental data are taken from Ku et
  al. \cite{Ku2012} (for $T/T_F < 2$) and from Sommer et
  al. \cite{Sommer} (for $T/T_F > 2$).
  \label{fig:kappa}}
\end{figure}

Although not related to the polarized Fermi gas, we can also study the
compressibility of the unpolarized gas above $T_c$. Following
\Ref{Sommer}, the compressibility is defined as $\kappa =
(\partial\rho/\partial\mu)/\rho^2$, where $\rho = \rho_\up =
\rho_\down$ is the density per spin state. Again, the computation of
$\kappa$ requires both the correlated density and the self-energy.
Our results for $\kappa$, normalized by the susceptibility of an ideal
Fermi gas at zero temperature, $\kappa_0(T=0) = 3/(2E_F\rho)$, are
shown in \Fig{fig:kappa}
together with the ideal Fermi-gas result $\kappa_0$ and the
experimental data from \Refs{Sommer,Ku2012}. We observe that the
compressibility is higher than that of the ideal Fermi gas, which is
plausible for a system with attractive interaction. For $T\gtrsim 0.8
T_F$, our results agree very well with the experimental ones. At lower
temperatures, the nice agreement is lost. When $T$ approaches $T_c$
from above, a strong increase of the compressibility is found in both
theory and experiment. However, since our $T_c$ is too high, we find
this increase at a higher temperature than observed in experiment.
\section{Summary and discussion}\label{sec:conclusions}
Although very successful in the description of the BCS-BEC cross-over
of an unpolarized two-component Fermi gas, the Nozi\`eres-Schmitt-Rink
(NSR) approach fails in the case of finite polarization $P$. As it was
already pointed out in previous work
\cite{LiuHu06,ParishMarchetti07,Kashimura2012}, the NSR scheme gives
in some regions of the phase diagram $\rho_\up < \rho_\down$ in spite
of $\mu_\up > \mu_\down$, and even in the unpolarized case the spin
susceptibility has the wrong (negative) sign. 

In this work we have suggested a way how to overcome the problems of
the original NSR approach in the polarized case. As in the NSR scheme,
we start from the ladder approximation for the T matrix and the
single-particle self-energy $\Sigma_\sigma$. We split $\Sigma_\sigma$
into a constant mean-field like shift $U_\sigma$ and an
energy-dependent part describing the correlations. While the
correlations are treated perturbatively, the shift $U_\sigma$ is
included self-consistently. This is different from the original NSR
approach, where the self-energy $\Sigma_\sigma$ as a whole is included
only to first order in the truncated Dyson series. We retrieve a
well-known formula for the correlation correction to the density,
originally derived by Zimmermann and Stolz (ZS) \cite{ZimmermannStolz}
for the unpolarized case.

Within the ZS scheme, the correlation correction to the density,
$\rho^{(1)}$, does not depend on the spin, which is plausible since
the correlated pairs are made of one atom of each spin. Another
interesting property of this approach is that $\rho^{(1)}$ vanishes in
the limit $T\to 0$, as required by the Luttinger theorem. Apparently,
by including the energy shift $U_\sigma$ one recovers in the $T\to 0$
limit the results of the particle-particle random-phase approximation
formulated within the $T=0$ formalism \cite{UrbanSchuck}.

When calculating occupation numbers, one finds that near unitarity and
at low temperature, the correlations become too strong to be treated
perturbatively. Apparently in these cases one cannot avoid to sum the
Dyson series to all orders, as it was done, e.g., in
\Refs{Perali2002,Kashimura2012}. This is beyond the scope of this work
and it is also not clear whether such a resummation would respect,
e.g., the equality of the correlation densities of both spins and the
Luttinger theorem in the $T\to 0$ limit. 

In the unpolarized case, the ZS approach interpolates, as the NSR
approach, between the BCS and BEC limits, however it reaches the BCS
limit much faster than the NSR approach. Near the unitary limit, the
critical temperatures for a given density are much lower than the BCS
one but still too high because screening effects of the
Gor'kov-Melik-Barkhudarov type \cite{GMB} are not included.

In contrast to the NSR scheme, the ZS scheme allows us to calculate
the phase diagram also as a function of polarization, since the
polarization has the same sign as the difference between the chemical
potentials, as it should. At not too strong polarizations, the
generalized ZS approach predicts a second-order phase transition
towards a BCS- or Sarma-like superfluid phase. At higher polarization
and lower temperature, one finds instead a transition towards a
FFLO-like phase where the Cooper pairs have a finite momentum. This is
in contrast to the experimental results obtained in the unitary limit
\cite{Shin2008} which show a first-order phase transition with
phase-separation between normal and superfluid phases at high
polarization and low temperature. In order to study a possible
first-order transition theoretically, one would need a theory that
describes both the normal and the superfluid phase. Another problem is
the critical polarization for the transition towards the FFLO phase at
$T=0$, which is much too high in our approach.

The spin susceptibility of the unpolarized gas within our approach is
positive, as it should be. It is smaller than that of an ideal Fermi
gas, which is also plausible. It agrees more or less with the
theoretical prediction of the extended T matrix approximation
(ETMA) \cite{Kashimura2012} and of the Luttinger-Ward theory
\cite{EnssHaussmann}, but not with the experimental results of
\Ref{Sommer}.

It should be noted that, even though the self-consistent energy shift
is a first step into that direction, one is still very far from a
fully self-consistent scheme such as that of
\Ref{EnssHaussmann,Haussmann} where all lines in \Fig{fig:feynman}
would correspond to dressed Green's functions. A less ambitious
improvement would be the so-called renormalized RPA
\cite{Catara,Delion} which amounts to replacing in \Eq{J} the Fermi
functions $f(\xi^*_{\kv\sigma})$ by the self-consistent occupation
numbers $n_{\kv\sigma}$. This would probably reduce the strong
correlations, especially at low temperature, and therefore help to
reduce the critical polarization of the FFLO phase.
\begin{acknowledgments}
We thank M. Zwierlein for sending us the data of \Ref{Ku2012}.
\end{acknowledgments}

\end{document}